\documentclass[12pt]{article}
\usepackage{graphicx}

\begin{document}

\title{Persistence and Success in the Attention Economy}
\author{Fang Wu and Bernardo A. Huberman\\HP Laboratories\\Palo Alto, CA 94304} \maketitle

\bigskip
\begin{abstract}
A hallmark of the attention economy is the competition for the attention of others. Thus people persistently upload content to social media sites, hoping for the highly unlikely outcome of topping the charts and reaching a wide audience. And yet, an analysis of the production histories and success dynamics of 10 million videos from \texttt{YouTube} revealed that the more frequently an individual uploads content the less likely it is that it will reach a success threshold. This paradoxical result is further compounded by the fact that the average quality of submissions does increase with the number of uploads, with the likelihood of success less than that of playing a lottery.
\end{abstract}
\pagebreak

Anyone trying to seek attention to a new idea, scientific result, or media creation faces the challenge of having it attended to by a sizable audience. This age old problem has become exacerbated with the advent of the web, which allows content to be easily published by millions of people while increasing the competition that it faces for the attention of users.\footnote{This is because information-rich regimes are characterized by keen competition for the user's attention \cite{falkinger-07,simon}.}  Worse, attention over content is distributed in a highly skewed fashion, implying that while a few items receive a lot of attention, most receive a negligible amount \cite{adamic-00,klamer-02,wu-07}. And yet people persistently upload content to social media sites, hoping for the highly unlikely outcome of topping the charts and reaching a wide audience. As in other endeavors, in the attention economy \cite{falkinger-07,franck-99,goldhaber-98,hirshleifer} persistence is construed by content producers as a key to success. And yet, an analysis of the production histories and success dynamics of 10 million videos from \texttt{YouTube} shows that the more frequently an individual uploads content the less likely it is that it will reach a success threshold. This paradoxical result is further compounded by the fact that the average quality of submissions does increase with the number of uploads, with the likelihood of success less than that of playing a lottery.

\texttt{YouTube} is the most popular video website on the Internet, where users can upload, view, and share video clips for free. In environments such as \texttt{YouTube}, the notion of success is not measured by monetary value but by the amount of attention each video receives \cite{huberman-09}. After a \texttt{YouTube} content producer uploads a video, a view count number is immediately displayed next to the video title, which measures how many times it has been watched. As the video is downloaded by more users, its view count grows accordingly. If a video has eventually been viewed a substantial number of times, it is promoted to \texttt{YouTube}'s front page, where it can reach a larger audience and receive even more attention.

\begin{figure}
\centering
\begin{minipage}{2.5in}
\centering\includegraphics[width=2.5in]{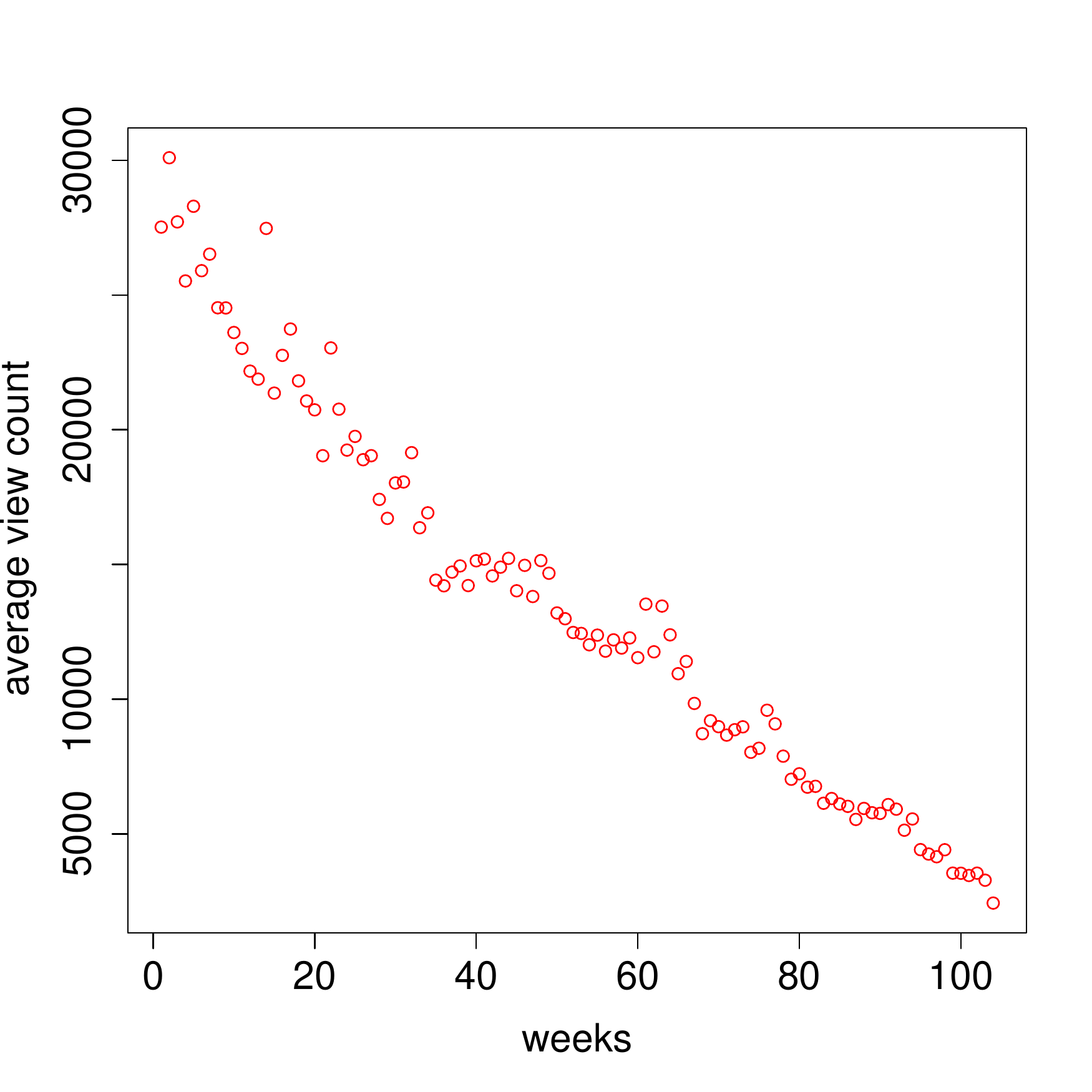}\\{\small (a)}
\end{minipage}
\begin{minipage}{2.5in}
\centering\includegraphics[width=2.5in]{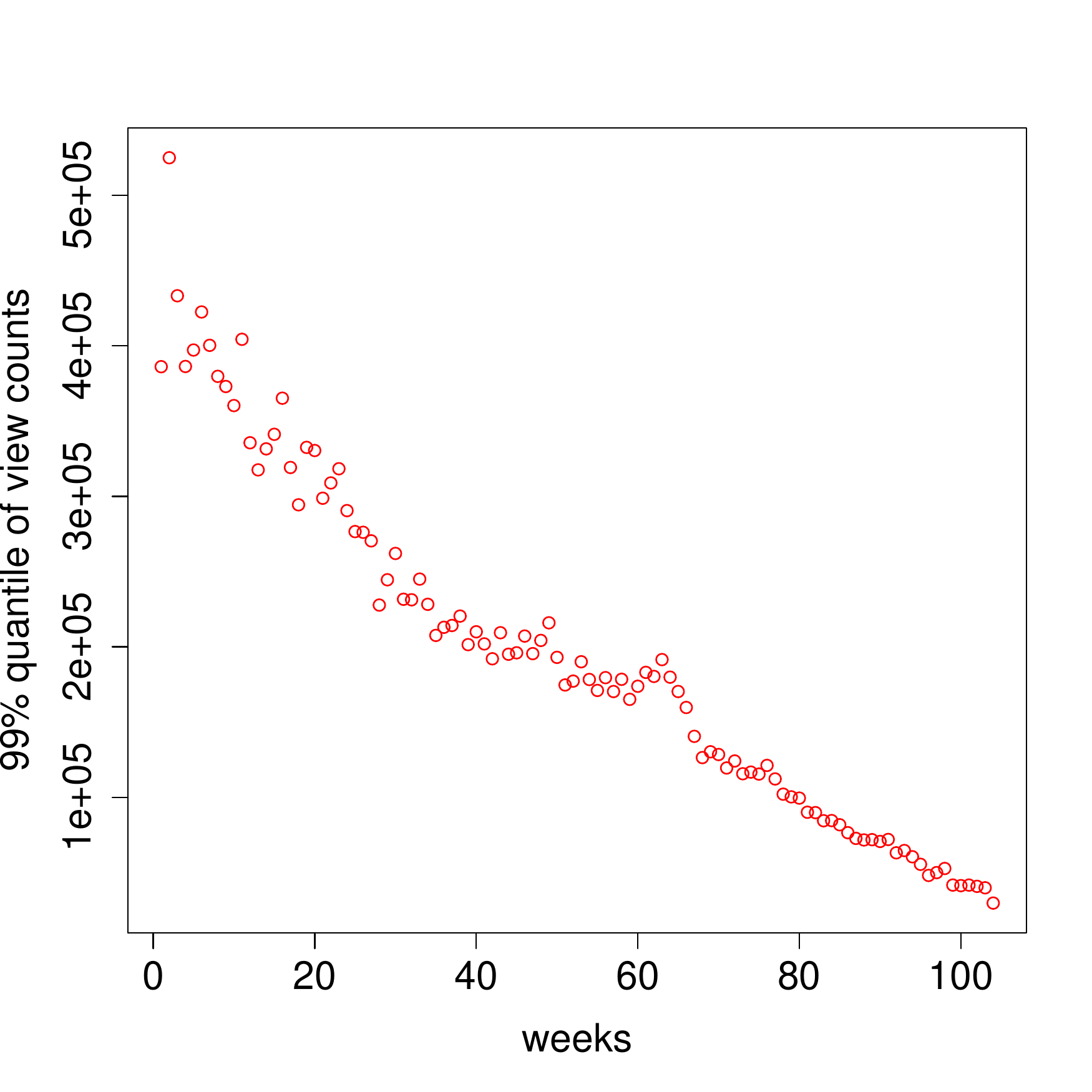}\\{\small (b)}
\end{minipage}
\caption{\label{fig:meanv}(a) Average view count of videos uploaded in different weeks. The horizontal axis represents the number of weeks from January 1, 2006. The vertical axis marks the average view count of videos uploaded in the same week. All videos' view counts are measured on April 30, 2008. As can be seen, older videos tend to receive more view counts than newer videos, because they have been on \texttt{YouTube} for a longer time. (b) 99\% quantile of videos uploaded in the same week, again exhibiting a declining trend.}
\end{figure}

Our data set contained 9,896,784 videos submitted by 579,470 producers by April 30, 2008. For each video we obtained its date stamp, the uploader's id, and its view count as of April 30, 2008. In what follows, we will say that a video is successful if its view count exceeds a popularity threshold, and unsuccessful if not. Obviously such a definition depends on the threshold that one chooses, as does any other definition of success. Also, because videos are uploaded at different times, and older videos tend to receive more view counts than newer videos, it only makes sense to compare the view count of videos released not far apart in time (Fig.~\ref{fig:meanv}). In what follows we will define a video to be a success if its view count is among the top 1\% of all videos uploaded in the same week, and call it a failure if otherwise. We also define a producer to be successful if at least one of her uploaded videos crosses the success threshold.

\begin{figure}
\centering
\begin{minipage}{2.5in}
\centering\includegraphics[width=2.5in]{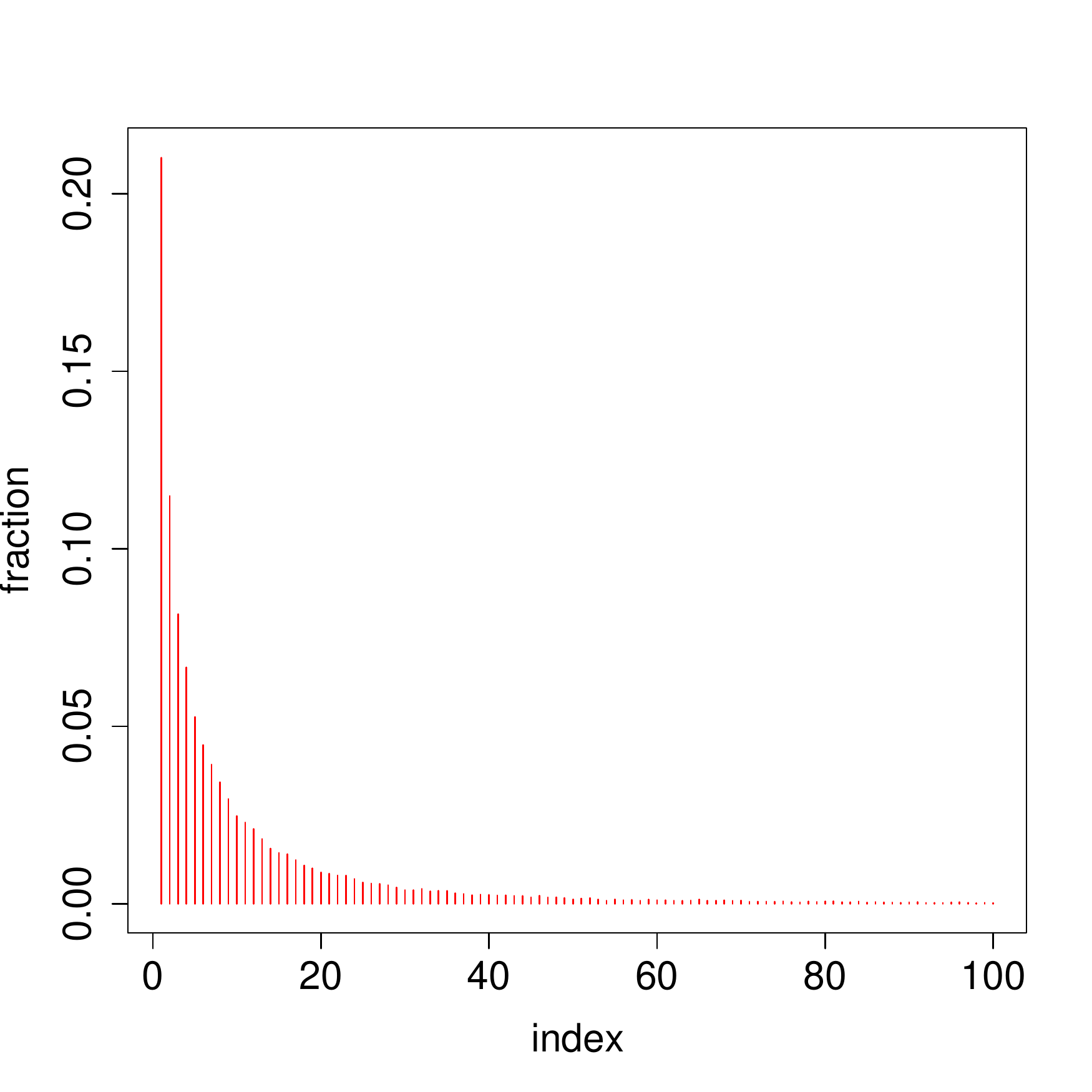}\\{\small (a)}
\end{minipage}
\begin{minipage}{2.5in}
\centering\includegraphics[width=2.5in]{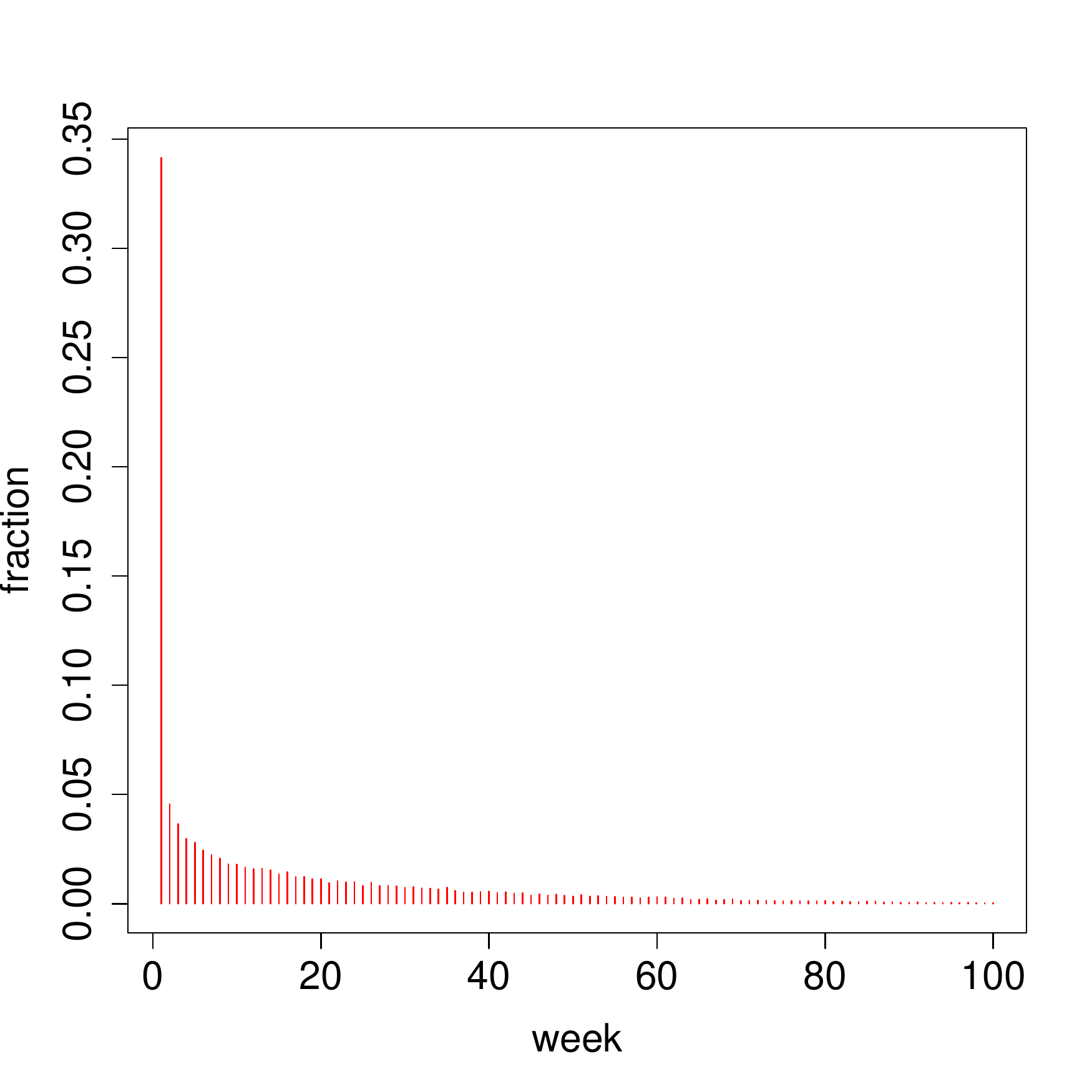}\\{\small (b)}
\end{minipage}
\caption{\label{fig:first success}(a) Fraction of successful producers who succeeded with their $k$'th video upload, where $k$ is the upload index. (b) Fraction of successful producers who succeeded in the $k$'th week since their first upload.}
\end{figure}

We start our analysis with some descriptive statistics. By definition 1\% videos in our data-set were successful. 7.36\% of the 579,470 successful producers, or 42,658 producers, uploaded at least one successful video before April 30, 2008 and hence belong to the set of successful producers. We observed that about 21.0\% of all successful users earned the success label with their first video. Given that only 5.9\% of all videos are first submissions, this number is impressively large. What is even more impressive is that about 34.1\% of the successful producers actually succeeded in their first week (i.e.~at least one video they uploaded in their first week received top 1\% popularity within that week). Fig.~\ref{fig:first success} plots the distribution of the index and week of the successful producers' first success. Again, one can see from the figure that a considerable proportion of producers succeeded early in their career, with only a few submissions within the first few weeks. In particular, more than half of the successful ones achieved success before uploading 5 videos, and more than half of those succeeded within 6 weeks.

While these observations seem to suggest that many producers enjoyed a swift success without having to be persistent, they do not imply that persistence is a bad strategy to achieve success. Persistence, by its very nature, enhances the probability of success simply because one is willing to endure more failures before achieving success. Thus, to answer the question of whether or not persistence is the key to success one needs to clarify the definition of both persistence and success.

According to the dictionary, the word ``persist'' is defined as ``to go on resolutely or stubbornly in spite of opposition, importunity, or warning.'' In our setting this corresponds to those producers who keep uploading despite the fact that their previous uploads never achieved the top 1\% popularity. Formally, let us define a producer's \emph{persistence level} to be the number of failures he/she is willing to endure before the first success. Thus, a user with persistence level $k$ ($k\ge 0$) would upload no less than $k$ videos before coming to a stop, and would refrain from uploading the $(k+1)$'th video if all of his/her first $k$ attempts fail. We also say that user A is more/less persistent than user B if A's persistence level is higher/lower than B.

Notice that there are two reasons why a producer's persistence level is not a directly observable quantity within our data-set. First, our data ends on April 30, 2008. If a producer uploaded $k$ videos before April 30, 2008 and all of them failed, we do not know whether he/she would still upload a $(k+1)$'th video after April 30, 2008. Second, if a producer failed $k$ times and then got a first success, all we can infer is that her persistence level must be no less than $k$. In fact, if a producer was lucky enough to have succeeded in the first video, we cannot make any inferences as to her persistence level.

\begin{figure}
\centering
\includegraphics[width=3in]{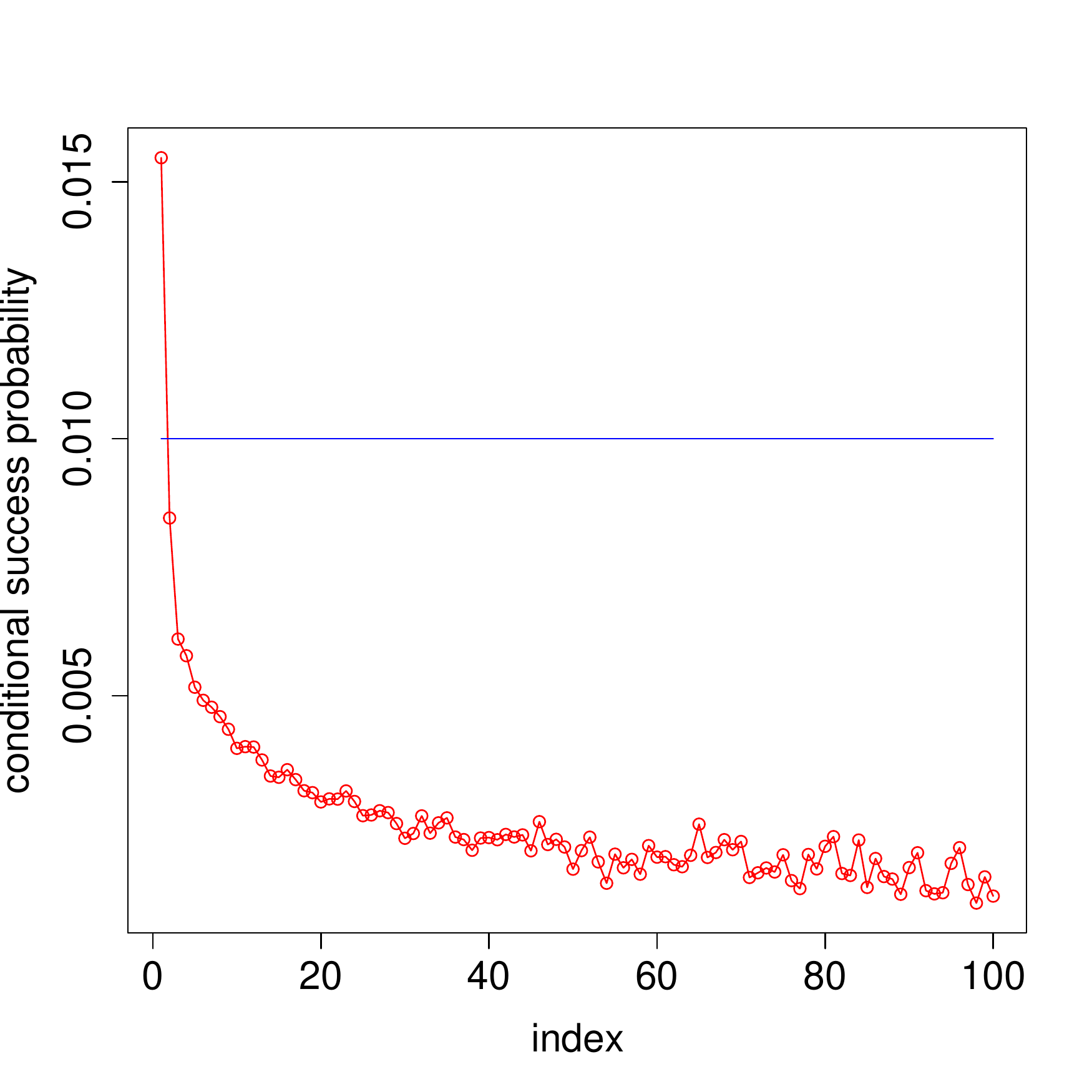}
\caption{\label{fig:hazard}The conditional success probability of \texttt{YouTube} producers. The blue line is the conditional success probability of a lottery with 0.01 winning probability.}
\end{figure}

Although it is not possible to measure the persistence level of each individual producer, one can measure the \emph{conditional success probability} or \emph{hazard function} of the whole population, which is defined as the conditional probability that the $k$'th attempt is a success given $k-1$ leading failures. For our data this amounts to
\begin{equation}
\label{eq:hazard}
h(k) = \frac{\parbox{3.5in}{number of producers who failed in the first $k-1$ videos and succeeded in the $k$'th video}}{\parbox{3.5in}{number of producers who failed in the first $k-1$ videos and still uploaded a $k$'th video}}.
\end{equation}
Fig.~\ref{fig:hazard} plots the conditional success probability for \texttt{YouTube} producers. We see that $h(k)$ declines with the submission index, indicating that later submissions are less likely to succeed. This result is somewhat surprising for one might expect the contributor to learn from past experiences and improve the chance to succeed. As the data shows, this is not the case.

A possible explanation of this surprising result could stem from the often implicit, but crucial assumption, that improvements along the way are what make persistence the cause of success.\footnote{For a counterexample see \cite{audia-00}.} Indeed, if the same mistake is repeated over and over again, what is the point of being persistent? In order to test whether success due to persistence with improvement is more than a matter of luck we compare our results with the benchmark case of a lottery with constant winning probability. While someone with no budget constraints can try to win a lottery as often as possible, the number of attempts will neither improve nor worsen the winning probability of the \emph{next} lottery purchase. In this sense the lottery can be regarded as a game of ``pure luck'', in which persistence does not play a role in determining winning success.

In order to compare our results from \texttt{YouTube} with the benchmark case of a lottery, we plot in Fig.~\ref{fig:hazard} the conditional success probability of a lottery with winning probability 0.01, where 0.01 is the fraction of successful videos that defines success in our study of \texttt{YouTube}. Naturally the conditional success probability for such a lottery is always 0.01, independent of past history.\footnote{The fact that the conditional success probability is a constant relies on the implicit assumption that all content providers have the same probability of success. Assuming heterogeneity of ability among the content providers will yield a decreasing function, but not one as steep as observed (unpublished).}  As can be seen in the figure, the conditional winning probability for \texttt{YouTube} is \emph{worse} than the lottery from the second video thereafter.

\begin{figure}
\centering
\includegraphics[width=3in]{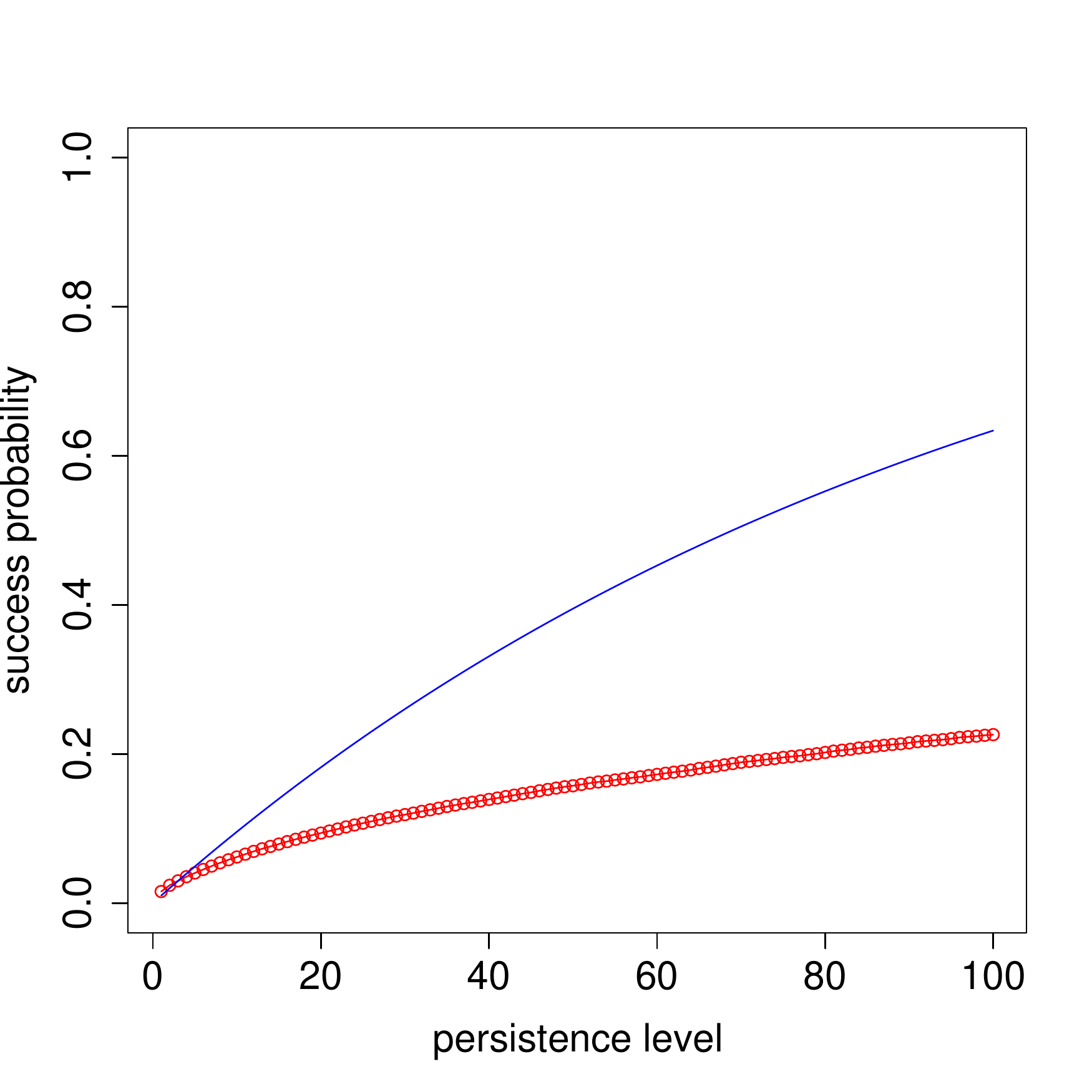}
\caption{\label{fig:success prob}The red line plots the success probability of \texttt{YouTube} producers as a function of their persistence level. The blue line plots the success probability of a producer who participates in a lottery with 0.01 winning probability for each draw, as a function of her persistence level.}
\end{figure}

More insight into this result can be obtained by calculating the success probability $p(k)$ of a producer with persistence level $k$, i.e.~the probability that the producer succeeds at least once before he/she quits. It is not hard to show that for the lottery case $p(k) = 1-0.99^k$, while for \texttt{YouTube} it is given by $p(k)= 1-\prod_{i=1}^k (1-h(i))$, where $h(k)$ is estimated by Eq.~(\ref{eq:hazard}). These two functions are plotted in Fig.~\ref{fig:success prob}. As can be seen, the success probability for a \texttt{YouTube} producer is lower than that of a lottery buyer regardless of the persistence level. In particular, while the success probability of a lottery buyer with persistence level 100 is 63.4\%, that of a \texttt{YouTube} producer with the same persistence level is only 22.6\%. Thus, the benefit of persistence on \texttt{YouTube} is much worse for content producers than that of a lottery.

\begin{figure}
\centering
\begin{minipage}{2.5in}
\centering\includegraphics[width=2.5in]{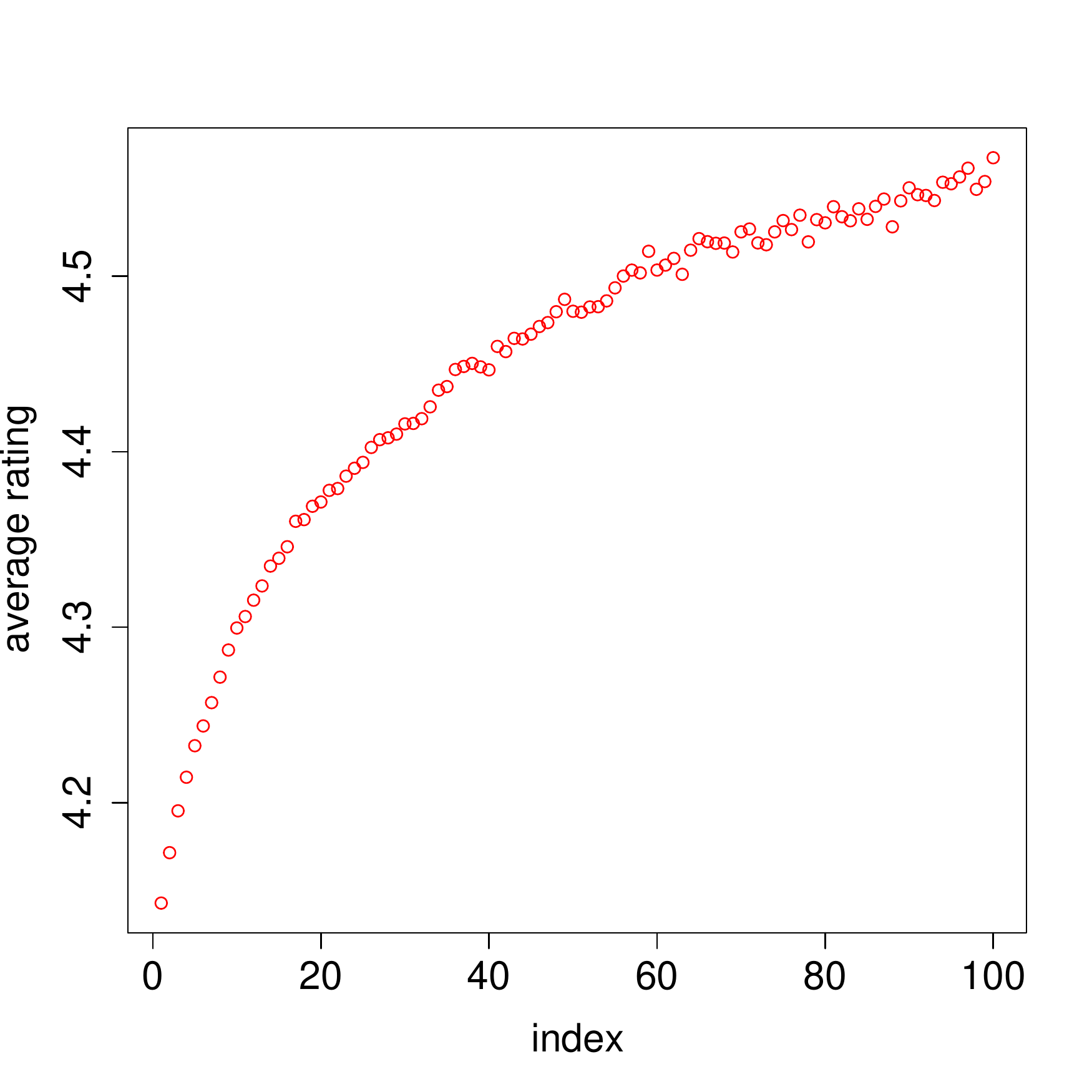}\\{\small (a)}
\end{minipage}\begin{minipage}{2.5in}
\centering\includegraphics[width=2.5in]{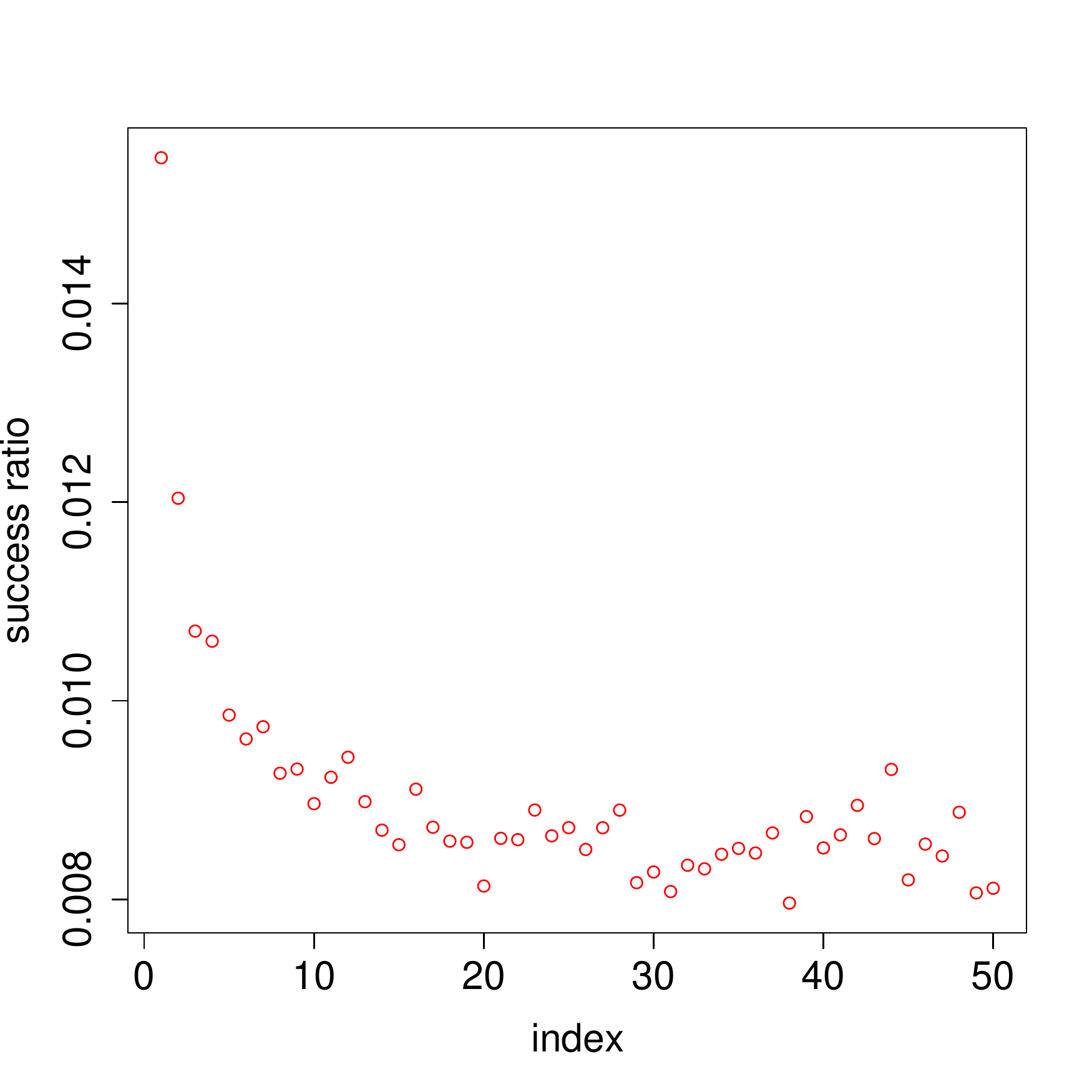}\\{\small (b)}
\end{minipage}
\caption{\label{fig:rating}(a) Average rating of videos with different submission index. A producer's first video has index 1, second video has index 2, and so on. (b) Average success ratio of videos with different submission index.}
\end{figure}

One question that remains to be answered is why it becomes harder for a \texttt{YouTube} producer to succeed as he/she submits another video. This question is especially puzzling when one considers the many learning advantages the producer could get from past submissions: improved video creation skills, better knowledge about the audiences' taste, better marketing techniques, etc. Fig.~\ref{fig:rating} illustrates this puzzle clearly. The producers on average receive higher ratings for their later videos, and yet their success ratio declines. If the video quality goes up, why should it receive less attention? This is particularly notable since the chances for success are independent of the ever increasing number of providers, given that the threshold for success is, by definition, fixed at a given percentage.

One possible explanation is that when a producer submits several videos over time, their novelty and hence their appeal to a wide audience tends to decrease. While this is one of many possible explanations we do not feel that we have a definitive answer to this paradox.

Finally there is the behavioral question of why people persist in uploading videos in light of the small odds of making it in the charts. A possible answer is that as in many other instances, individuals overestimate the true probabilities of winning when they are small, thus increasing their efforts while their chances of becoming successful decrease over time. This is similar to the longshot anomalies observed and discussed in the context of gambling \cite{thaler-88}. While in those cases plausible explanations have been suggested, we do not have enough behavioral data from content providers to elucidate their motivations.

In summary, we have shown that in a pure attention economy the more frequently an individual uploads content the less likely it is that it will reach a success threshold. In contrast to the extensive literature on attention at the individual level \cite{kahneman-73}, this is a collective effect rather than a single individual one, since the success we measured is a competitive outcome. Furthermore, this paradoxical result is compounded by the fact that the average quality of submissions does increase with the number of uploads, with the likelihood of success less than that of playing a lottery. These results throw light into a pure form of the attention economy which is close to the theoretical idealizations studied in recent years \cite{falkinger-07}.

\end{document}